# An electron as a wave packet
# in the relativistic strophotron fel

M. Hnatic[1,2,3] , P. Kopcansky [1], K.B. Oganesyan [4,*]

* bsk@yerphi.am

[1] Institute of Experimental Physics SAS, Kosice, Slovakia

[2] Joint Institute for Nuclear Research, Dubna, Russia

[3] Faculty of Sciences, P. J. Safarik University, Kosice, Slovakia

[4] A. Alikhanyan National Lab, Yerevan Physics Institute, Yerevan, Armenia

We study the initial distribution of electrons in relativistic strophotron FEL over vibrational levels determined by the expansion coefficients. We show the quantity of the vibrational level $l_0$ can be expressed in terms of the initial parameters of the electron beam.

Free-Electron Lasers [1-4] are powerful, tunable, coherent sources of radiation, which are used in scientific research, plasma heating, condensed matter physics, atomic, molecular and optical physics, biophysics, biochemistry, biomedicine etc. FELs today produce radiation ranging from millimeter wavelengths to X rays, including parts of spectrum in which no other intense, tunable sources are available [5]. This field of modern science is interesting from the point of view of fundamental research and very promising for further applications.

There are many alternative schemes of free electron lasers: Cherenkov, transition radiation, Smith-Purcell, ubitron, orotron types and so on (see [1-5] and references therein.]). One of the alternative schemes is strophotron fel.

The classical and quantum theories of free electron lasers of strophotron (SFEL) type are developed in [6-40[. It is shown, that there is a strong inhomogeneous broadening (depending on electron transverse coordinate at entrance in strophotron) of spectral distribution of spontaneous radiation and the gain in SFEL, which was not taken into account in [6]. This broadening brings to overlapping of spectral lines of spontaneous radiation and the gain.

In the present paper we will bring correspondence between quantum and classical approaches for SFEL theories using wave packet presentation of electrons.

We shall consider an ultrarelativistic electron of energy $\varepsilon = m\gamma >> m$ ($\hbar = c = 1$) moving in the *XOZ* plane at a small angle $\alpha$ to the *OZ* axis and experiencing at an initial moment t=0 a static electric or magnetic field with the potential or

$$\Phi(x) = \Phi_0(x^2 / d^2) \qquad \mathbf{A}(x) = \mathbf{A}_0(x^2 / d^2), \qquad (1)$$

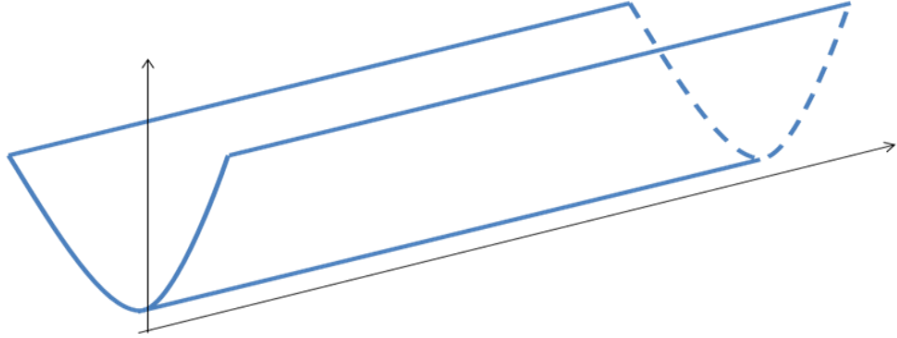

where $e\Phi_0 = e|\mathbf{A}_0|$ and 2d are, respectively, the maximum height of a transverse potential barrier and the size of the region of its localization (aperture) in the direction of the *OX* axis; the $\mathbf{A}_0$ vector is parallel to the *OZ* axis, i.e., the magnetic field is directed along the *OY* axis.

In the absence of an electromagnetic wave an electron experiences translational motion along the *OZ* axis, characterized by a momentum $\mathbf{p}_{\parallel}$ and an energy $\varepsilon_{\parallel} = \sqrt{p_{\parallel}^2 + m^2}$, oscillating in the transverse direction (along the *OX* axis). The oscillation wave functions $\varphi_l(x, \varepsilon_{\parallel})$, satisfy the usual Schrodinger equation obtained from the initial Klein-Gordon equation ignoring the square of the potential $\Phi(x)$) [21,39]:

$$d^2\varphi_l(x, \varepsilon_{\parallel}) / dx^2 = \left[ 2\varepsilon_{\parallel} e\Phi(x) - 2\varepsilon_{\parallel}\Omega(\varepsilon_{\parallel})(l + 1/2) \right] \varphi_l(x, \varepsilon_{\parallel}), \qquad (2)$$

where $\Omega(\varepsilon_{\parallel}) = \left( 2e\Phi_0 / \varepsilon_{\parallel} \right)^{1/2} / d$ is the oscillation frequency in a parabolic well; l= 0, 1,2,... are the oscillation quantum numbers; $\varphi_l(x, \varepsilon_{\parallel}) = \left[ \Omega(\varepsilon_{\parallel})\varepsilon_{\parallel} \right]^{1/4} \chi_l(\varsigma)$, $\chi_l(\varsigma)$ are the usual oscillation functions dependent on the dimensionless variable $\varsigma = (\Omega\varepsilon_{\parallel})^{1/2}$ and expressed in terms of Hermite polynomials [43].

Formula

$$\omega \approx \omega_{res}^{(s)} = 2(2s+1)\gamma^2\Omega / (1 + l_0\Omega\gamma^2 / \varepsilon) \equiv (2s+1)\omega_{res}, \qquad (3)$$

with

$$\omega_{res} = 2\gamma^2\Omega / (1 + l_0\Omega\gamma^2 / \varepsilon),$$

determines the resonant frequency $\omega_{res}$ of the system, expressed in terms of the number of the vibrational level $l_0$ most efficiently populated at the initial moment of time $t = 0$.

The quantity $l_0$ can be expressed in terms of the initial parameters of the electron beam. If $\psi^{(0)}(x)$ is an initial transverse electron function, then the initial distribution over vibrational levels is determined by the expansion coefficients $\psi^{(0)}(x)$ in $\varphi_l(x)$

$$c_l = \int \psi^{(0)}(x)\varphi_l(x)dx \qquad (4)$$

The function $\psi^{(0)}(x)$ should be specified in the form of a packet localized at the point $x_0$, where $x_0$ is the initial value of the electron transverse coordinate. For convenience, we choose $\psi^{(0)}(x)$ in the form of a Gaussian function

$$\psi^{(0)}(x) = \left(\frac{2}{\pi d_q^2}\right)^{1/4} \exp\left\{ip_\perp(x - x_0) - \frac{(x - x_0)^2}{d_q^2}\right\} \qquad (5)$$

where $p_\perp$ is the initial transverse impulse, $d_q$ is the quantum "smearing" of an individual electron ($1/d_q$ is the packet width).

Let us calculate the integral $\psi^{(0)}(x)$ (4). Substituting into it explicit expressions for (5) and the vibrational wave function $\varphi_l(x)$ $\varphi_l(x, \varepsilon_\parallel) = \left[\Omega(\varepsilon_\parallel)\varepsilon_\parallel\right]^{1/4}\chi_l(\varsigma)$, $\chi_l(\varsigma)$ are the usual oscillation functions dependent on the dimensionless variable $\varsigma = (\Omega\varepsilon_\parallel)^{1/2}$ and expressed in terms of Hermite polynomials [43] and using the well-known tabular integral for the Hermite polynomials $H_l(\xi)$ [44]

$$\int d\xi e^{-(\xi - \eta)^2} H_l(\alpha\xi) = \sqrt{\pi}\left(1 - \alpha^2\right)^{\frac{l}{2}} H_l\left(\frac{\alpha y}{\left(1 - \alpha^2\right)^{\frac{1}{2}}}\right)$$

find

$$c_l = \pi^{1/4}\left(\frac{2\sqrt{b}}{b + 1}\right)^{1/2}\left(\frac{b - 1}{b + 1}\right)^{1/2}\chi_l(\varsigma)\exp\left[\frac{a^2 b + 2iab^2\zeta_0 + b\zeta_0^2}{}\right] \qquad (6)$$

where

$$\zeta = -\frac{b\zeta_0 + ia}{\sqrt{b^2 - 1}}, \quad \zeta_0 = \sqrt{\Omega\varepsilon} \equiv \frac{x_0}{d_0}$$

and the dimensionless parameters a and b are introduced: $a = p_\perp / \sqrt{\Omega\varepsilon}$ is the ratio of the size of the localization region $d_0 = 1/\sqrt{\Omega\varepsilon}$ of the ground vibrational state $\chi_0(x/d_0)$ to the transverse de Broglie wavelength $\hbar_{DB} = 1/p_\perp$ of the electron (almost $d_0 >> \hbar_{DB}$ always a >> 1), $b = 2/(\Omega\varepsilon d_q^2) = 2\left(d_0/d_q\right)^2$ is the double square of the ratio $d_0$ to the packet width $d_q$ (5).

For large $l$ , for the vibrational function $\chi_l(\zeta)$ , we use the semiclassical solution of the corresponding Schrödinger equation

$$\chi_l(\zeta) = \sqrt{\frac{2}{\pi}} \frac{1}{\left(2l - \zeta^2\right)^{1/4}} \cos\left\{ \int_\zeta^{\sqrt{2l}} d\zeta \sqrt{2l - \zeta^2} - \frac{\pi}{4} \right\} \tag{7}$$

Note, that the well-known asymptotic formulas for Hermite polynomials with a large index [43] are valid for $2l >> \zeta_0^2$ and in this approximation they follow from (7). For our purposes, these formulas are insufficient, since in the region of the maximum distribution $|c_l|^2$ can be $l \sim \zeta^2$ (see below).

The pre-exponential factor in formula (7) cannot be too large, because near the classical turning point $\zeta^2 = 2l$, and the condition for the applicability of the quasiclassics

$$\left| \zeta^2 - 2l \right| >> (2l)^{2/3} \tag{8}$$

is violated .

In the most important cases, the condition (8) considered below is satisfied in the region of the maximum of the distribution (8) With this in mind, we further investigate only the main exponential dependence of $c_l$ on, $l$ omitting the pre exponential factor in (7).

Since the scale of the quantum smearing of an electron at the exit from the accelerator is unknown, we will consider separately two possible cases $b >> 1$ ($d_q << d_0$), and $b << 1$ ($d_q >> d_0$).

1. $b >> 1, d_q << d_0$ . The factor $[(b-1)/(b+1)]^{1/2}$ , in equation (6), takes the form $\exp(-l/b)$ in this case. The argument $\zeta$ of the vibrational function $\chi_l(\zeta)$ in (6) becomes equal to $\zeta = -\zeta_0 - i\frac{a}{b}$. The corresponding simplifications of expressions (6), (7) give

$$|c_l|^2 \sim \exp\left\{ \frac{2}{b} \left[ -\left( l - \frac{\zeta_0^2}{2} \right) + a\sqrt{2l - \zeta_0^2} \right] \right\} \tag{9}$$

Index of exponent (9) is maximum at

$$l = l_0 \equiv \frac{1}{2}\left(a^2 + \zeta_0^2\right) = \frac{1}{2}\left(\frac{p_\perp^2}{2\varepsilon} + x_0^2\Omega\varepsilon\right) = \frac{\varepsilon}{2\Omega}\left(\alpha^2 + x_0^2\Omega^2\right) = \frac{\varepsilon\Omega a^2}{2} \qquad (10)$$

where $\alpha$ is the entrance angle of electron in the system.

In the surrounding area

$$|c_l|^2 \sim \exp\left\{-\frac{(l - l_0)^2}{4a^2 b}\right\} \qquad (11)$$

The positions of the maximum $|c_l|^2$ (9), $l_0$ (10) are determined both by the angle of entry of electrons into the strophotron $\alpha$ and by the initial value of the transverse coordinate $x_0$ of the electron. The distribution width $|c_l|^2$ of (11) is

$$\delta l = 2a\sqrt{b} = \frac{\sqrt{2}p_\perp}{\Omega\varepsilon d_q} = \frac{\sqrt{2}d_0^2}{\lambda_{D.B._\perp} d_q} \qquad (12)$$

It is easy to see, that $\dfrac{\delta l}{l_0} \leq \lambda_{\textit{ДБ}\perp} / d_{\textit{кв}}$. If the quantum "smearing" of the electron is such, that $\lambda_{\textit{ДБ}\perp} \ll d_{\textit{кв}} \ll d_0$, the distribution $|c_l|^2$ (9), (11) is narrow ($\delta l \ll l_0$). With growth of $d_{\textit{кв}}$, width $\delta l$ (12) decreases. Note, that under typical conditions $\lambda_{\textit{ДБ}\perp} \sim 3 \cdot 10^{-11} \textit{см}$, $d_0 \sim 3 \cdot 10^{-7} \textit{см}$.

In the case considered, the distance between the points $l = l_0$ and $l = l_c = \frac{1}{2}\zeta^2 \approx \frac{1}{2}\zeta_c^2$ is equal to $\frac{1}{2}a^2 \sim l_0 \gg \delta l, l_0^{2/3}$, i.e. the maximum of the distribution $|c_l|^2$ is located far from the classical turning point $l_c$, and over the entire width $\delta l$ of the distribution $|c_l|^2$, the applicability of the (8) quasiclassical system is not violated.

2. $b \ll 1$, $d_q \gg d_0$. In this case, the factor $[(b-1)/(b+1)]^{l/2}$, in formula (6), takes the form $(-1)^{l/2}\exp(-bl)$, and argument $\zeta$ in the function $\chi_l(\zeta)$ becomes equal to

$$\zeta = sign\left(a - ib\zeta_0\right) = \frac{p_\perp signx_0}{\sqrt{\Omega\varepsilon}} - \frac{2i|x_0|}{\sqrt{\Omega\varepsilon}d_q^2} \qquad (13)$$

The sign function at (13) corresponds to the larger of the two exponential terms that contribute to the sum of (7) $\cos[...] = \frac{1}{2}\left(e^{il[...]} + e^{-il[...]}\right)$ в (7).

Simplification of expression (7) for $l > \dfrac{a^2}{2}$ gives

$$|c_l|^2 \sim \exp\left\{-2b\left[l - \frac{a^2}{2} - |\zeta_0|\right]\sqrt{2l - a^2}\right\} \qquad (14)$$

The index of exponent (14) is maximum at the previous value of $l_0$ (10). Expanding the exponent (14) in the vicinity of the maximum, obtain

$$|c_l|^2 \sim \exp\left\{-\frac{b(l - l_0)^2}{\zeta_0^{\,2}}\right\} \qquad (15)$$

The distribution width in this case is

$$\delta l = \frac{\zeta_0}{\sqrt{b}} = \sqrt{2}\,\Omega\varepsilon\,|x_0|\,d_q \qquad (16)$$

Expansion in (15) on $l - l_0$ is justified, if $\delta l$ is small compared to the distance between $l_0$ and the classical turning point $l_c = \dfrac{1}{2}\zeta^2 \approx \dfrac{1}{2}a^2$, and, if $l_0 - l_c > l_0^{2/3}$ (8) i.e. if $|x_0| >> d_{\kappa\epsilon}$ and $x_0^{\,2}\Omega\varepsilon >> (p_\perp^{\,2}/\Omega\varepsilon)^{2/3}$. Under typical conditions. $x_0 \sim d_e \sim \alpha/\Omega$, $l_0 >> 1$ these conditions are satisfied.

In this case $\delta l << l_0$, till $d_q << d_e$, i.e. the quantum smearing of the electron is much less than the transverse size of the beam. Apparently, this condition is also always met.

Thus, in both considered cases $|c_l|^2$ has a maximum at $l = l_0$ (10), and the distribution width is determined by expressions (12) and (16) at $d_q << d_0$ and $d_q >> d_0$.

Under very natural and realistic assumptions of $d_q$ the distribution $|c_l|^2$ is narrow $\delta l << l_0$.

## Summary

It is shown, that the distribution function $|c_l|^2$ is narrow in both real considered cases and has maximum at $l_0 = \dfrac{\varepsilon}{2\Omega}\left(\alpha^2 + x_0^{\,2}\Omega^2\right)$, which verifies assumption made in [21, 39], i.e. dependence of system resonance frequency, spectral distribution of spontaneous radiation and the gain on initial parameters of electron.

## Acknowledgements

KBO thanks NSP of the Slovak Republic for support..

## References

1. J.M.J. Madey, J. Appl. Phys. 42, 1906 (1971)
2. L.R. Elias, J.M.J. Madey, H.A. Schwettman, T.I. Smith, Phys. Rev. Lett., 36, 717, 1976.
3. C.A. Brau. Free-Electron Lasers, Boston, Academic, 1990.
4. M.V. Fedorov. Atomic and Free Electrons in a Strong Light Field, Singapore, World Scientific, 1997.
5. https://accelconf.web.cern.ch/fel2019/papers/proceed.pdf. Proceedings of FEL'19, Hamburg, Germany, 2019.
6. D. F. Zaretsky, E. A. Nersesov," *Zh. Eksp. Teor. Fiz.*, **84**, pp. 892-902, 1983.
7. M.V. Fedorov, K.B. Oganesyan, IEEE Journal of Quant. Electr, 21, 1059 (1985).
8. K.B. Oganesyan, J. Mod. Optics, 62, 933 (2015).
9. Fedorov M.V., Oganesyan K.B., Prokhorov A.M., Appl. Phys. Lett., 53, 353 (1988).
10. Oganesyan K.B., Prokhorov A.M., Fedorov M.V., Sov. Phys. JETP, 68, 1342 (1988).
11. Oganesyan KB, Prokhorov AM, Fedorov MV, Zh. Eksp. Teor. Fiz., 53, 80 (1988).
12. K.B. Oganesyan, M.L. Petrosyan, YerPHI-475(18) – 81, Yerevan, (1981).
13. Petrosyan M.L., Gabrielyan L.A., Nazaryan Yu.R., Tovmasyan G.Kh., Oganesyan K.B., Laser Physics, 17, 1077 (2007).
14. E.A. Nersesov, K.B. Oganesyan, M.V. Fedorov, Zhurnal Tekhnicheskoi Fiziki, 56, 2402 (1986).
15. K.B. Oganesyan, J. Mod. Optics, 61, 763 (2014).
16. D.N. Klochkov, AI Artemiev, KB Oganesyan, YV Rostovtsev, CK Hu, J. of Modern Optics, 57, 2060 (2010).
17. K.B. Oganesyan. Laser Physics Letters, 12, 116002 (2015).
18. GA Amatuni, AS Gevorkyan, AA Hakobyan, KB Oganesyan, et al, Laser Physics, 18, 608 (2008).
19. K.B. Oganesyan, J. Mod. Optics, 62, 933 (2015).
20. A.I. Artem'ev, DN Klochkov, K Oganesyan, YV Rostovtsev, MV Fedorov, Laser Physics 17, 1213 (2007).
21. Zaretsky, D.F., Nersesov, E.A., Oganesyan, K.B., Fedorov, M.V., Sov. J. Quantum Electronics, 16, 448 (1986).
22. K.B. Oganesyan, J. Contemp. Phys. (Armenian Academy of Sciences), 50, 123 (2015).
23. DN Klochkov, AH Gevorgyan, NSh Izmailian, KB Oganesyan, J. Contemp. Phys., 51, 237 (2016).
24. K.B. Oganesyan, M.L. Petrosyan, M.V. Fedorov, A.I. Artemiev, Y.V. Rostovtsev, M.O. Scully, G. Kurizki, C.-K. Hu, Physica Scripta, T 140, 014058 (2010).
25. Oganesyan K.B., Prokhorov, A.M., Fedorov, M.V., ZhETF, 94, 80 (1988).
26. D.N. Klochkov, AI Artemiev, KB Oganesyan, YV Rostovtsev, MO Scully, CK Hu, Physica Scripta, T 140, 014049 (2010).
27. K.B. Oganesyan, J. Contemp. Phys. (Armenian Academy of Sciences), 52, 91 (2017).

28. A.I. Artemyev, M.V. Fedorov, A.S. Gevorkyan, N.Sh. Izmailyan, R.V. Karapetyan, A.A. Akopyan, K.B. Oganesyan, Yu.V. Rostovtsev, M.O. Scully, G. Kuritzki, J. Mod. Optics, 56, 2148 (2009).

29. A.S. Gevorkyan, K.B. Oganesyan, Y.V. Rostovtsev, G. Kurizki, Laser Physics Lett., 12, 076002 (2015).

30. K.B. Oganesyan, J. Contemp. Phys. (Armenian Academy of Sciences), 50, 312 (2015).

31. K.B. Oganesyan, J. Contemp. Phys. (Armenian Academy of Sciences), 50, 123 (2015).

32. DN Klochkov, AH Gevorgyan, NSh Izmailian, KB Oganesyan, J. Contemp. Phys., 51, 237 (2016).

33. 79. K.B. Oganesyan, M.L. Petrosyan, M.V. Fedorov, A.I. Artemiev, Y.V. Rostovtsev, M.O. Scully, G. Kurizki, C.-K. Hu, Physica Scripta, T 140, 014058 (2010).

34. Oganesyan K.B., Prokhorov, A.M., Fedorov, M.V., ZhETF, 94, 80 (1988).

35. K.B. Oganesyan. Laser Physics Letters, 13, 056001 (2016).

36. DN Klochkov, KB Oganesyan, YV Rostovtsev, G Kurizki, Laser Physics Letters 11, 125001 (2014).

37. K.B. Oganesyan, Nucl. Instrum. Methods A 812, 33 (2016).

38. M.V. Fedorov, K.B. Oganesyan, IEEE Journal of Quant. Electr, 21, 1059 (1985).

39. D.F. Zaretsky, E.A. Nersesov, K.B. Oganesyan, M.V. Fedorov, Kvantovaya Elektron. 13 685 (1986).

40. K.B. Oganesyan, J. Mod. Optics, 61, 1398 (2014).

41. K.B. Oganesyan, J. Contemp. Phys. (Armenian Academy of Sciences), 51, 307 (2016).

42. K.B. Oganesyan, Journal of Contemporary Physics (Armenian Academy of Sciences) 51, 10 (2016).

43. A. S. Davydov, Quantum Mechanics, Pergamon Press, Oxford (1965).

44. I.S. Gradshteyn and I.M. Ryzhik, Table of Integrals, Series, and Products, Seventh ed., Academic Press, 2007.